\documentclass[prd,twocolumn,showpacs,nofootinbibs
]{revtex4}
\usepackage{epsf}
\usepackage{amsmath}

\def\beq{\begin{equation}}
\def\eeq{\end{equation}}
\def\bea{\begin{eqnarray}}
\def\eea{\end{eqnarray}}
\def\beqa{\begin{equation}\begin{array}{l}}
\def\eeqa{\end{array}\end{equation}}
\def\eqlab#1{\label{eq:#1}}
\def\figlab#1{\label{fig:#1}}

\def\eref#1{(\ref{eq:#1})}
\def\Eqref#1{Eq.~(\ref{eq:#1})}

\def\Figref#1{Fig.~\ref{fig:#1}}

\def\barr{\left(\begin{array}{c}}
\def\earr{\end{array}\right)}
\def\bmat{\left(\begin{array}{cc}}
\def\emat{\end{array}\right)}
\def\al{\alpha}

\def\ga{\gamma} 
 \def\De{\Delta}\def\vDe{\varDelta}

\def\w{\omega}

\def\pa{\partial}

\def\pa{\partial}

\def\nn{\nonumber}

\def\lag{{\mathcal L}}

\def\mathscr{\mathcal}

\def\3d{3-D}

\def\ol#1{\overline{#1}}

\begin{document}
\preprint{WM-06-111}
\preprint{JLAB-THY-06-581}

\title{Large-$N_c$ relations for the electromagnetic $N$ to 
$\Delta$(1232) transition}

\author{Vladimir Pascalutsa}
\email{vlad@ect.it}

\affiliation{European Centre for Theoretical Studies in Nuclear Physics and Related Areas (ECT*), Villa Tambosi, Villazzano
I-38050 TN, Italy}

\author{Marc Vanderhaeghen}
\email{marcvdh@jlab.org}
\affiliation{Physics Department, College of William and Mary,
Williamsburg, VA 23187, USA}
\affiliation{Theory Center, Thomas Jefferson National Accelerator Facility, 
Newport News, VA 23606, USA}

\date{\today}

\begin{abstract}
We examine the large-$N_c$ relations which express the 
electromagnetic $N$-to-$\Delta$  
transition quantities
in terms of the electromagnetic properties of the nucleon. 
These relations are based on the known large-$N_c$ relation
between the $N\to \De$ electric quadrupole moment and the
neutron charge radius, and a newly derived 
large-$N_c$ relation between the electric quadrupole ($E2$) and
Coulomb quadrupole ($C2$) transitions. Extending these relations
to finite, but small, momentum transfer we find that the description 
of the electromagnetic $N\to\Delta$ ratios ($R_{EM}$ and $R_{SM}$)
in terms of the nucleon form factors predicts a structure which
may be ascribed to the effect of the ``pion cloud''. 
These relations also provide useful 
constraints for the $N \to \Delta$ generalized parton distributions. 
\end{abstract}


\maketitle
\thispagestyle{empty}

\section{Introduction}

The electromagnetic properties of the nucleon, such as magnetic moments and charge radii,
provide a benchmark information on the nucleon structure.
In recent years the nucleon electromagnetic 
form factors (FFs) have been charted very 
precisely, thanks to the new generation of 
experiments that make use of the target- and recoil-polarization techniques, 
see \cite{HydeWright:2004gh,Arrington:2006zm,Perdrisat:2006hj} for recent reviews. 
These precise data allow, e.g., to map out the spatial 
densities in the nucleon, address the role of the meson 
cloud, study the transition to the asymptotic regime.

The $N \to \Delta$ FFs, describing the electromagnetic 
transition of the nucleon to its first excited state, contain 
complementary information, such as the sensitivity on the nucleon 
shape, see e.g., \cite{Papanicolas:2007zz,Pascalutsa:2006up}.
It is therefore interesting to see that, in the limit of a large number of 
colors ($N_c$) \cite{'t Hooft:1973jz,Witten:1979kh}, QCD provides relations between the nucleon and $N \to \Delta$ properties. For example, the
relation between the isovector 
nucleon magnetic moment and the $N \to \Delta$ transition magnetic 
moment ($M1$)  
is well known~\cite{Jenkins:1994md}. In this paper we establish 
a large-$N_c$ relation between the small electric quadrupole ($E2$) and 
Coulomb quadrupole ($C2$) $N \to \Delta$ amplitudes, and relate them to 
the neutron charge radius, see Sect.~II. Furthermore, we extend these relations to finite  
momentum transfer ($Q^2$) and hence express the $N \to \Delta$ 
FFs in terms of the nucleon FFs, see Sect.~III. 
We shall then use a recent empirical parameterization of the nucleon FFs 
to test the large-$N_c$ relations on the ratios $E2/M1$ and $C2/M1$,
for which precise experimental data are available as well.
The main points of this study are summarized in Sect.~IV.

\section{Large-$N_c$ relations}

To introduce the electromagnetic $N\to \De$ transition 
we start with an effective $\ga N\De$
Lagrangian~(see, e.g,~\cite{Pascalutsa:2003zk,Pascalutsa:2005vq}):
\bea
\eqlab{lagran}
\lag_{\ga N\De} &=&   \frac{3 i e}{2M_N (M_N + M_\Delta)} \,\ol N \, T^3 \nn\\
&\times & \left[  g_M \,\pa_{\mu}\De_\nu\, \tilde F^{\mu\nu}  
 + i g_E \,\ga_5\,\pa_{\mu}\De_\nu \, F^{\mu\nu} \right. \\
&- & \left.  \,\frac{g_C}{M_\De} \ga_5 \ga^\al  
(\pa_{\al}\De_\nu-\pa_\nu\De_\al) \,\pa_\mu F^{\mu\nu}\right] + \mbox{H.c.},
\;\;\;\;\; \nn
\eea
where $N$ denotes the nucleon (spinor) 
and $\De_\mu$  the $\De$-isobar (vector-spinor) fields, 
$M_N$ and $M_\De$ are respectively their masses, $F^{\mu\nu}$ and $\tilde F^{\mu\nu}$
are the electromagnetic field strength and its dual, 
$T^3$ is the isospin-1/2-to-3/2 transition operator. 
An important observation here is that the couplings $g_M$, $g_E$ and $g_C$ appear with the same
structure of spin-isospin and field operators, and hence we expect them to
scale with the same power of $N_c$, for large $N_c$.

It is customary to characterize the three different 
types of the $\ga N \De$ transition
in terms of the Jones--Scadron FFs~\cite{Jones:1972ky}: $G^*_M$,  $G^*_E$,
$G^*_C$. The contribution of the effective couplings entering \Eqref{lagran} 
to these FFs can 
straightforwardly be computed, with the following result:
\begin{eqnarray}
\label{eq:JS}
G_M^\ast(Q^2) &=& g_M \,+\left( -M_\De \,\w\,g_E
+  Q^2 g_C\right)/Q_+^2\,\,, \nn\\ 
G_E^\ast(Q^2) &=& \left(-M_\De \,\w\, g_E+  Q^2  g_C\right)/Q_+^2\,\,, \\
G_C^\ast(Q^2) &=&-2M_\De\, \left( \w \,  g_C + M_\De \,g_E\right)/Q_+^2\,\,,\nn
\end{eqnarray}
where $\w$ is the 
photon energy in the $\Delta$ rest frame: $\w=(M_\De^2-M_N^2-Q^2)/(2M_\De)$
and we use the notation: 
\beq
Q_\pm = \sqrt{(M_\De\pm M_N)^2 +Q^2}\,.
\eeq 
At $Q^2 = 0$, we immediately find:
\begin{subequations}
\bea
\eqlab{GEstar}
G_E^\ast(0) &=& - \frac{\vDe}{2 (M_N+M_\De)}  \, g_E , \\
G_C^\ast(0) &=& - \frac{2M_\De^2}{(M_N+M_\De)^2} 
\left[ \frac{M_\De^2-M_N^2 }{2M_\De^2 } \,  g_C +  g_E\right]
\eqlab{GCstar}
\eea 
\end{subequations}
where $\vDe \equiv M_\De - M_N$ is the $\De$-nucleon mass difference. 
In the large-$N_c$ limit this mass difference goes as $1/N_c$, 
whereas the baryon masses increase proportionally to $N_c$:
\beq
M_{N (\De)} = {\mathcal O}(N_c),\,\,\, \vDe = {\mathcal O}(N_c^{-1}) \,.
\eeq
Given the fact that, for large $N_c$, $g_E$ and $g_C$ scale with the same power of $N_c$, 
the first term in \Eqref{GCstar} is suppressed by $1/N_c^2$, and we obtain
the following  relation:
\beq
G_C^\ast(0) \,=\, \frac{2 M_\Delta}{M_N+M_\Delta} \,\frac{2M_\De}{\vDe} \,G_E^\ast(0)\,.
\label{eq:gc0largenc}
\eeq

Of special interest are the multipole ratios: $R_{EM}=E2/M1$ and $R_{SM}=C2/M1$,
which in terms of the Jones-Scadron FFs are given by:
\beq
\eqlab{ratios}
R_{EM} = - \frac{G_E^\ast}{G_M^\ast} \,,\quad \quad 
R_{SM} = - \frac{Q_+ Q_-}{4M_\De^2} \, \frac{G_C^\ast}{G_M^\ast}.
\eeq
It is easy to see that the relation \eref{gc0largenc} 
tells us that these ratios are equal ($R_{SM} = R_{EM}$) for large $N_c$ and $Q^2=0$.
Note also that using Eqs.~\eref{GEstar} and \eref{ratios}
one easily recovers the result of Ref.~\cite{Jenkins:2002rj}: $R_{EM}={\mathcal O}(1/N_c^2)$. 

Let us now recall the other relevant large-$N_c$ results. Namely, 
the magnetic $N\to\De$ transition FF is related to the isovector
anomalous magnetic moment of the nucleon, $\kappa_V\simeq 3.7$, 
as~\cite{Jenkins:1994md}:
\beq
\eqlab{GMlargeNc}
G_M^\ast(0) = \frac{1}{\sqrt{2}} \, \kappa_V\,, 
\eeq
whereas the $N \to \Delta$ 
quadrupole moment $Q_{p \to \Delta^+}$ 
can be related to the neutron charge radius $r_n$ as~\cite{Buchmann:2002mm}:
\begin{eqnarray} 
Q_{p \to \Delta^+} \,=\, \frac{1}{\sqrt{2}} \, r_n^2 .
\label{eq:qpdelrnlargenc2}
\end{eqnarray}
The latter relation can directly be expressed in terms of the $E2$ Jones--Scadron FF:
\footnote{The precise relation between the two $E2$ quantities is
$$ G_E^\ast(0) = -\mbox{$\frac{1}{12}$}
(M_N/M_\De)^{3/2}\, (M_\De^2 -M_N^2) \,Q_{p \to \Delta^+}\,.$$
}
\begin{eqnarray}
G_E^\ast(0) \,=\, - \,
\frac{M_\Delta^2 - M_N^2}{12\sqrt{2}}\,\left(\frac{M_N}{M_\De}\right)^{3/2}\! r_n^2 .
\label{eq:ge0largenc}
\end{eqnarray}
Moreover, using the new relation \eref{gc0largenc}, we can express the 
$C2$ $N\to\De$ FF in terms of the neutron charge radius as well:
\beq
G_C^\ast(0) \,=\, - \,\sqrt{2M_N M_\De} \, M_N\, r_n^2/6\,.
\eeq

Therefore,  the $\ga\, N\to \De$ transition ratios in the large-$N_c$ limit can be
expressed entirely in terms of the nucleon electromagnetic properties:
\beq
\eqlab{central}
R_{EM} = R_{SM} = 
\frac{1}{12}\,\left(\frac{M_N}{M_\De}\right)^{3/2}\!(M_\Delta^2 - M_N^2)\,
\frac{r_n^2}{\kappa_V}.
\eeq
This is one of the central results of this work.\footnote{A similar set of relations for
the $\ga N\De$ ratios was derived earlier by Buchmann~\cite{Buchmann:2004ia} 
using a combination
of the large-$N_c$ expansion and the spin-flavor symmetry. Our results rely
solely on the large-$N_c$ limit.}

Empirically, the large-$N_c$ limit value for $M1$ is off
by about 15\%; compare $G_M^\ast(0) = \kappa_V / \sqrt{2} \simeq
2.62 $ with  the empirical value~\cite{Tiator:2003xr}: 
$G_M^{*}(0) \simeq 3.02$. The value for $E2$, 
$G_E^\ast(0) \simeq 0.07$, obtained from \Eqref{ge0largenc}
using the experimental neutron charge radius~\cite{Sick:2005az}
 ($r_n^2 = -0.113\pm 0.003 $~fm$^2$), is in a better agreement with
the empirical value~\cite{Tiator:2003xr}: $G_E^\ast(0) \simeq 0.075$. The strength of $C2$,
$G_C^\ast(0) \simeq 0.7$ is also somewhat smaller than the empirical
value~\cite{Pascalutsa:2006up,Pascalutsa:2005vq}: $G_C^\ast(0) \simeq 1.0\,$.

For the ratios we
then obtain:
\beq
\eqlab{numcentral}
R_{EM} = R_{SM} = - 2.77 \, \%\,.
\eeq
For $R_{EM}$ this large-$N_c$ 
prediction is in an excellent agrement with experiment~\cite{PDG2006}:
$R_{EM} = - 2.5 \pm 0.5 \%$. For $R_{SM}$, a direct measurement at the real-photon
point is of course not possible. In the following section we will examine an 
extension of the large-$N_c$ to finite $Q^2$, which will in particular 
allow for a direct comparison with the experimental data for $R_{SM}$.

\section{Extension to finite momentum transfer}

It would be desirable to extend the above relations to finite $Q^2$.
For example, the most straightforward generalization of 
\Eqref{GMlargeNc} gives~\cite{Frankfurt:1999xe}:   
\beq
G_M^{*}(Q^2) 
=  \frac{1}{\sqrt{2}} 
\left\{ F_{2p}(Q^2) - F_{2n} (Q^2) \right\} \, ,  
\label{eq:gmsumrule} 
\eeq
where $F_{2p} - F_{2n}$ is the (isovector) combination of the 
proton (p) -- neutron (n) Pauli FFs. 

Analogously, extending the newly derived relation~\Eqref{gc0largenc}
to finite, but small $Q^2$,  we have
\begin{eqnarray}
G_C^\ast (Q^2) = \frac{4 M_\Delta^2}{M_\Delta^2 - M_N^2} \, 
G_E^\ast(Q^2),
\label{eq:gcqlargenc}
\end{eqnarray} 
or, equivalently, for the ratios \Eqref{ratios}:
\begin{eqnarray}
R_{SM}(Q^2) = \frac{Q_+ \, Q_-}{M_\Delta^2 - M_N^2} \, 
R_{EM}(Q^2) .
\label{eq:rsmlargenc}
\end{eqnarray}
Finally, we may use the fact that, for small $Q^2$, 
the neutron electric FF is expressed  
as $G_{En}(Q^2) \approx - r_n^2 \, Q^2 / 6$, and hence an extension 
of \Eqref{ge0largenc} is given by:
\begin{eqnarray}
G_E^\ast(Q^2) = 
\left(\frac{M_N}{M_\De}\right)^{3/2} \frac{M_\Delta^2 - M_N^2}{2\sqrt{2}\,Q^2} \, 
G_{En}(Q^2).
\label{eq:geqlargenc}
\end{eqnarray}

Bringing these results together we obtain the following 
expression for the $\ga N\De$ ratios in terms of the nucleon
form factors:
\begin{subequations}
\eqlab{RatiosLargeNc}
\bea
R_{EM} &= & - \left(\frac{M_N}{M_\De}\right)^{3/2}
\frac{M_\Delta^2 - M_N^2}{2Q^2}\frac{G_{En}}{F_{2p} - F_{2n}} \, 
, \\ 
R_{SM} &= & - \left(\frac{M_N}{M_\De}\right)^{3/2}
\frac{Q_+\,Q_-}{2Q^2}\frac{G_{En}}{F_{2p} - F_{2n}} \,. 
\eea
\end{subequations}

\begin{figure}
\centerline{  \epsfxsize=9cm%
  \epsffile{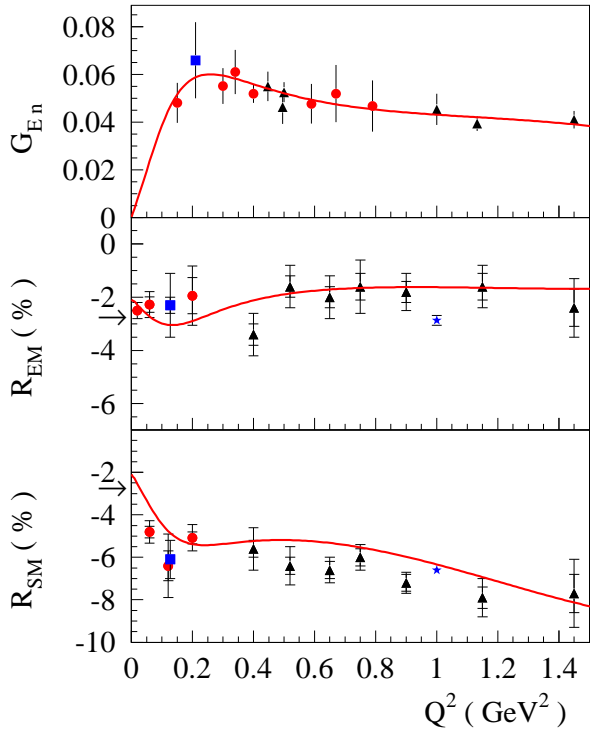} 
}
\caption{(Color online)  
Neutron electric FF, $G_{En}$, (upper panel) in comparison with  
the $N \to \Delta$ $R_{EM}$ (middle panel) 
and $R_{SM}$ (lower panel) ratios. The curve for $G_{En}$ shows
the empirical parameterization
of Bradford {\it et al.}\ \cite{Bradford:2006yz}. The curves
for $R_{EM}$  and $R_{SM}$ are obtained using 
the relations of \Eqref{RatiosLargeNc}
with the empirical nucleon FFs~\cite{Bradford:2006yz}.  
The data for $G_{E n}$ are from 
double-polarization experiments at 
MAMI~\cite{herberg,ostrick,becker,rohe,glazier} (red circles), 
NIKHEF~\cite{passchier} (blue square),  
and JLab~\cite{zhu,madey,warren} (black triangles). 
The data for $R_{EM}$ and $R_{SM}$ 
are from BATES~\cite{Sparveris:2004jn} (blue squares),   
MAMI~\cite{Beck:1999ge,Stave:2006ea,Sparveris:2006} (red circles), 
JLab/CLAS~\cite{Joo:2001tw} (black triangles),  
and JLab/HallA~\cite{Kelly05} (blue stars). 
}
\figlab{fig:remrsm}
\end{figure}

The latter  relations 
are tested in \Figref{fig:remrsm}, where we show the 
$Q^2$ dependence of the neutron electric FF 
and the resulting $N\to \De$ transition ratios, compared to experimental data.
The curves are obtained by using the empirical parameterization of 
the nucleon FFs by Bradford {\it et al.}~\cite{Bradford:2006yz},
together with the relations of \Eqref{RatiosLargeNc}
for $R_{EM}$ and $R_{SM}$.

The arrows on the $y$-axis indicate the large $N_c$ 
prediction \Eqref{numcentral}. 
The numerical values at $Q^2 = 0$ are slightly 
different from the ones obtained through \Eqref{RatiosLargeNc}
(solid curves) only because the form-factor 
parameterization of Bradford {\it et al.},
which is used here, does not exactly reproduce the neutron charge radius. 

We can see that the prediction of the $Q^2$ dependence for both 
ratios is in a very 
good agreement with the experimental data, even at higher momentum
transfers. At low $Q^2$, it is very interesting to see that 
the slight bump in $G_{En}$ around $Q^2 \simeq 0.2 $~GeV$^2$ 
results in a 
shoulder structure in both of the ratios, which seems to be consistent
with the experimental data. Friedrich and Walcher, 
in their model analysis of the nucleon FF data~\cite{Friedrich:2003iz},
observe such bump structures in all four nucleon electromagnetic 
FFs and attribute them to the effects of the ``pion cloud''. 
The present large-$N_c$ relations show that analogous effects must then  
arise in $R_{EM}$ and $R_{SM}$ at low $Q^2$. 
Although the available data for $R_{EM}$ and $R_{SM}$ 
seem to support this finding, it would certainly be necessary 
to improve on the accuracy of the data in the region of 
$Q^2 \simeq 0.10 - 0.25$~GeV$^2$, to provide a convincing 
evidence for such structures.  

We emphasize again that the relations of \Eqref{RatiosLargeNc}
are derived assuming that the momentum transfer is small, $Q^2\ll 1$ GeV$^2$.
Nevertheless, it is intriguing to see that their phenomenological
success extends into the region of intermediate $Q^2$, as is explored 
in \Figref{fig:remrsm}.

It is perhaps useful to point out that the excellent agreement of the
large-$N_c$ relations with experimental data for these quantities makes
it interesting to study the quark-loop effects (which are suppressed
in the large-$N_c$ limit~\cite{Witten:1979kh}) 
by comparing the quenched versus full QCD calculations 
for these quantities. Presently, quenched lattice QCD calculations
for $R_{EM}$ and $R_{SM}$ are done at larger than physical 
pion masses~\cite{Alexandrou:2004xn},
but indeed they compare favorably with
chiral perturbation theory predictions~\cite{Pascalutsa:2005ts}. 
It would be interesting
to see if this agreement persists for pion masses down to the
physical point.

\section{Conclusion}
In the large-$N_c$ limit of QCD, the properties of the
$N$ to $\De$ transition can be related to the properties
of the nucleon. We have shown here how 
the $E2$ and $C2$ $\gamma N \Delta$ 
transitions are related to each other and to the neutron charge radius
[\Eqref{central}]. 
We have extended these relations to low momentum transfers, thus relating 
the $R_{EM}$ and $R_{SM}$ ratios to the ratio of the 
neutron electric form factor over the nucleon isovector Pauli form factor
[\Eqref{RatiosLargeNc}].   

Using an empirical representation of the nucleon form factors, we have
tested the prediction of the large-$N_c$ relations for $R_{EM}$ and
$R_{SM}$ ratios versus the available experimental data.
The predictions for the ratios show a remarkable consistency with 
experiment. We note however that the 
predictions for the absolute strength of the transitions are 
less successful phenomenologically, the large-$N_c$ relations
underestimate the $M1$ and $C2$ strength by about 20\%. Evidently, the 
relations for the ratios work much better. 
 
It is particularly interesting to see that 
the large-$N_c$ relations translate
the structures in the nucleon form factors, 
which arise due to the long-range effects, into 
a dip structure in $R_{EM}$ and a shoulder structure in $R_{SM}$ around 
$Q^2 \simeq 0.15 - 0.25$~GeV$^2$. This prediction calls for more
precise measurements in that $Q^2$ range to confirm that such structures
are indeed present.
Finally, we remark that even at higher momentum transfers 
the large-$N_c$ relations for the ratios are in
a surprisingly good agreement with experiment, while  
the perturbative QCD prediction~\cite{Carlson:1985mm}
(i.e., $R_{EM} \to +100 \%$ and $R_{SM}\to \mbox{const}$,
as $Q^2\to \infty$) 
is nowhere in sight.
 
The large-$N_c$ relations between the nucleon and $N \to \Delta$
electromagnetic form factors can also be used to 
constrain the first moment of the
$N \to \Delta$ generalized parton distributions (GPDs), 
see \cite{Pascalutsa:2006up} for more details.
The relations examined here are relevant for the 
$E2$ and $C2$ $N\to \De$ GPDs, and may 
shed light on the quark distributions inducing the 
$\Delta$-resonance excitation.  

\begin{acknowledgments} 
This work is partially supported  by the European Community Research Infrastructure Activity under the FP6 "Structuring the European Research Area" programme (HadronPhysics, contract RII3-CT-2004-506078),
and in part by the U.S.\ Department of Energy grant no.\
DE-FG02-04ER41302 and contract DE-AC05-06OR23177 under
which Jefferson Science Associates operates the Jefferson Laboratory. 
\end{acknowledgments}

\end{document}